\documentclass[aps,twocolumn,showpacs]{revtex4}
\usepackage{graphicx}
\usepackage{epsf}
\usepackage{amsmath}
\usepackage{textcomp}
\usepackage{amssymb}

\begin{document}

\title{Optomechanics of random media}
\vspace{0.5cm}
\author{S. Gentilini$^{1,\star}$ and C. Conti$^{1,2}$}
\affiliation{\small
$^1$ ISC-CNR c/o Dipartimento di Fisica - Universit\`{a} La Sapienza, P. A. Moro
2, 00185, Roma, Italy\\
$^2$ Dipartimento di Fisica - Universit\`{a} La Sapienza, P. A. Moro
2, 00185, Roma, Italy\\
$^*$Corresponding author: silvia.gentilini@roma1.infn.it}

\date{\today}

\begin{abstract}
{\bf
Using light to control the movement of nano-structured objects is a great challenge.\cite{padgett14}
This challenge involves fields like optical tweezing  \cite{Ashkin78,dholakia:10,dileonardo:13}, Casimir forces \cite{Capasso}, integrated optics \cite{Arcizet06,Gigan06,Garbos:11}, bio-physics \cite{Russell2013,rdl13}, and many others.
Photonic ``robots'' could have uncountable applications. However, if the complexity of light-activated devices increases, structural disorder unavoidably occurs
and, correspondingly, light scattering, diffusion and localization. Are optically-driven mechanical forces affected by disorder-induced effects? A possible hypothesis is that light scattering reduces the optomechanical interaction. Conversely, we show that disorder is a mechanism that radically enhances the mechanical effect of light. We determine the link between optical pressure and the light diffusion coefficient, and unveil that when the Thouless conductivity becomes smaller than the unity, at the so-called Anderson transition, optical forces and their statistical fluctuations reach a maximum. Recent advances in photonics demonstrate the possibility of harnessing disorder for fundamental physics and applications \cite{issue13}. Designing randomness allows new materials with innovative optical properties \cite{leonetti11,mosk07,bertolotti12,Sebbah12,Folli13,ghofraniha13,cao13}.
Here we show that disorder and related phenomena may be exploited for optomechanical devices.
}
\end{abstract}

\maketitle
Several recent investigations show that light propagation and amplification in disordered matter can be controlled, and this allows applications as tunable random lasers, transmission through random media, and novel disorder-driven devices as random lasers and ultra-sensible spectrometers \cite{leonetti11, mosk07,bertolotti12,Sebbah12,Folli13,ghofraniha13,cao13}.
Controlling the photon-transport also enables to tailor the optical properties of new materials \cite{garcia07,Noh11}, albeit effects like the three-dimensional (3D) Anderson localization of light are still largely debated. \cite{john87,wiersma97,skipetrov14}
However, there is a specific field in which the benefits of randomness have not yet been investigated, and that is optomechanics.
Even if the concept of optical pressure can be dated back to the $16$th century, with the famous work of Keplero on comet tails \cite{kepler},
the role of disorder-induced light diffusion and localization on the optomechanical forces (OMFs) is unexplored.

OMF can be calculated by a field-dependent stress tensor \cite{MaxwellBook}, and in the last century many efforts were devoted to experimental and theoretical analyses \cite{Lebedev,Nichols} of the momentum transferred to a dielectric body by electromagnetic (EM) radiation, \cite{Minkowski08,Abraham09,Einstein17,Boyd2010} and to an enormous number of applications including, among others, laser cooling \cite{Arcizet06,Gigan06}, optical manipulation \cite{padgett14, Ashkin78,dholakia:10,dileonardo:13}, biophysics \cite{Russell2013,rdl13} and optomechanical devices \cite{Arcizet06,Gigan06,Garbos:11,butsch2012}.

OMFs are due to the interaction of the EM field with the boundaries of dielectric objects. When considering random systems with a large number of interfaces, determining the momentum transferred from photons to a material can be highly non trivial.
In the perspective of optically activated nano-structured devices, understanding the OMF due to many interfaces is pivotal for possible applications.

As described elsewhere \cite{balazs53,Boyd14} any photon that is transmitted unaltered through a dielectric material does not furnish kinetic momentum to matter, however, if the scattering changes the photon direction, a recoil force appears. In the presence of multiple scattering, the photon random walk generates a random walk of the medium.
This problem is remarkably similar to the random walk of a macroscopic object in a liquid: \cite{einstein05} even
if the timescale of the single molecule collision is very fast,
the statistical fluctuations of the number of collisions generate a slow observable motion.
For the photon, we find that the temporal fluctuations of the forces occur on a timescale much longer that the optical carrier period,
and that the physics is made even richer by the onset of localization effects.
Indeed, it is well accepted that when the photon transport mean free path $\ell$ is small enough, three-dimensional (3D) disordered systems can support long-living localized states \cite{Storzer06,Gentilini09,genack2011,maret2013}.
We investigate here the role of the localized states on the OMFs.

Due to the small values of the OMFs, one must consider dielectric systems with spatial dimensions of the order of tens of wavelengths. Larger system are not substantially affected by the OMFs. We hence analyze regimes in which EM localized states are strongly altered by finite-size effects, and this calls for fully-vectorial solutions of Maxwell equation. In the considered cases, the diffusion approximation (only meaningful for large systems) cannot be applied.

In the following, we hence apply a massively parallel computational approach to provide answers to the open question on the role of disorder
on the OMFs occurring in 3D random assembly of dielectric particles. By the Maxwell stress tensor method (MSTM), we find the relation between the OMFs exerted on the entire disordered system and the parameters characterizing the photons transport regime, as specifically the diffusion constant and the Thouless conductivity \cite{ShengBook}.
%--------------Figure 1------------------------%
\begin{figure*}
\includegraphics[width=16cm]{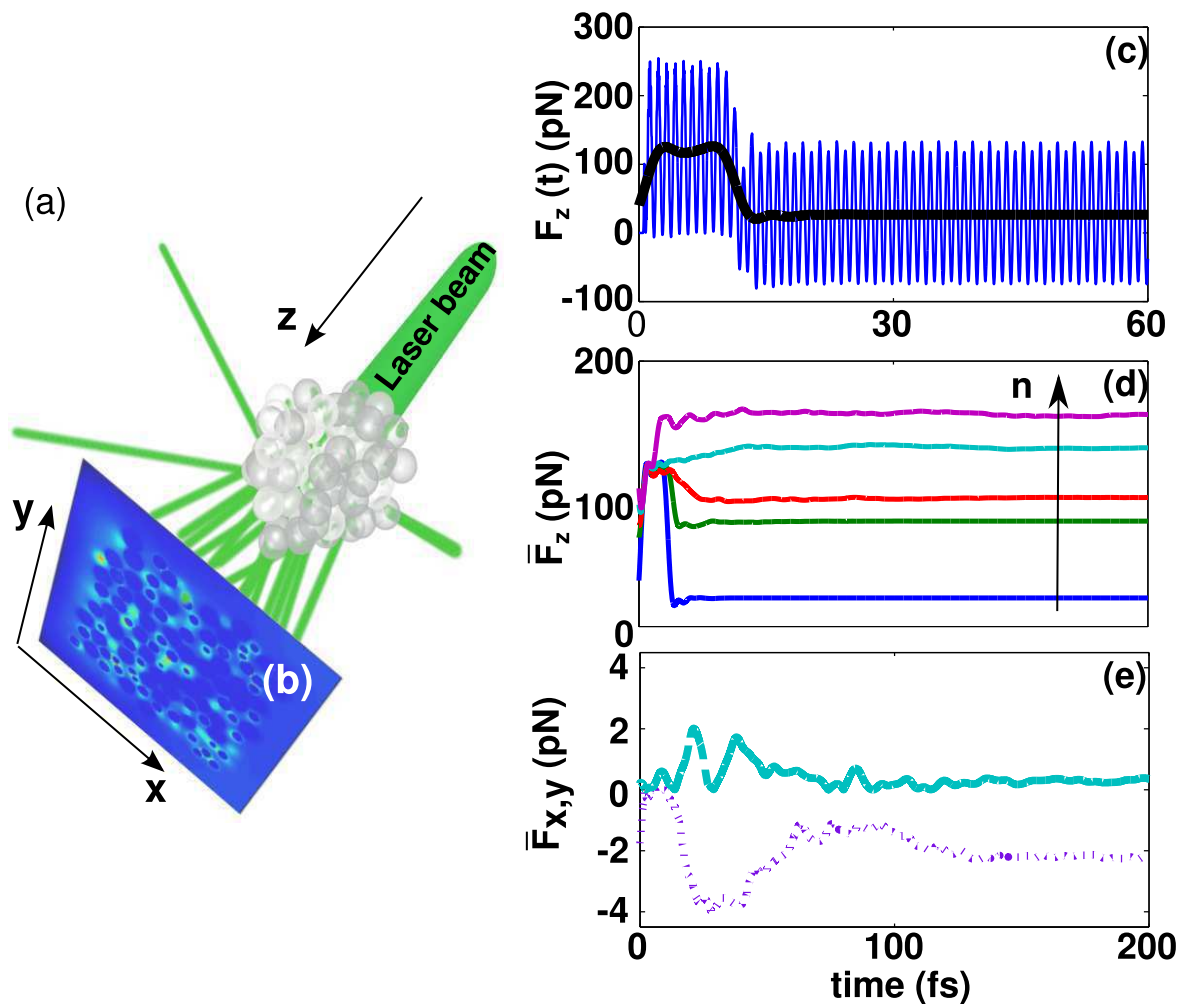}
\caption{\textbf{Numerical experiment.} (a) Sketch of the physical system considered in the 3D-FDTD numerical simulations; (b) output $(x,y)$ distribution of the electric field; (c) time-dynamics of the OMF $z$-component
$F_z(t)$: the fast oscillations are due to the optical cycle of the light source; the superimposed thick curve is obtained by spectral filtering; (d) time dynamics of $\overline F_z(t)$ for different particles refractive
index $n$ and for $L=2.0\mu$m; (e) as in (d) for the transverse forces $\overline F_x(t)$ (dotted line) and $\overline F_y(t)$ (continuous line) for $L=2.0\mu$m and $n=2.5$.}\label{fig1}
\end{figure*}
%----------------end figure 1-----------------%

\textbf{Numerical simulations of Maxwell equations.} Our numerical approach is based on a finite-difference-time-domain (FDTD) algorithm \cite{TafloveBook} with typical runs involving thousands of processors in an IBM Blue Gene/Q system with a massively parallel architecture. Other authors have previously calculated by FDTD techniques the optical pressure on dielectric media \cite{mahajan:03,Zhang:04,Gauthier:05,Jiang:06,Sung:08}; to the best of our knowledge disorder has not been analyzed before.
%--------------Figure 2------------------------%
\begin{figure*}
\includegraphics[width=16cm]{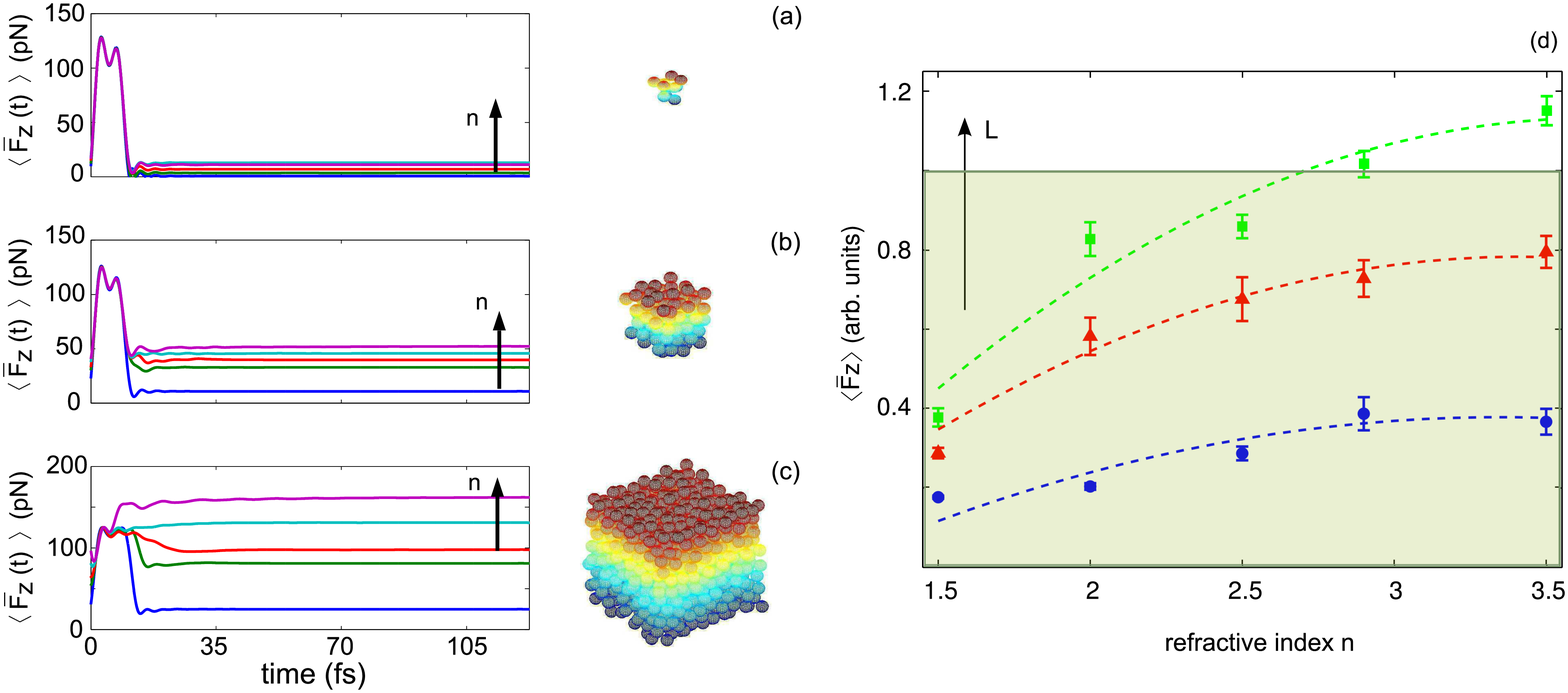}
\caption{\textbf{Disorder averaged behavior of the z-component OMF Vs scattering strength and system size.} (a-c) Disorder averaged  $\langle\overline F_z(t)\rangle$ for the different system sizes $L=0.5\mu$m (a), $1.5\mu$m
(b) and $2.0\mu$m (c). Each curve corresponds to a different particle refractive index $n=1.5$, $2.0$, $2.5$, $2.9$ and $3.5$ (from bottom to top). The 3D structures in the panels are representatives of the simulated assemblies.
(d) Stationary value of the curves reported in panels (a-c), $\langle \overline F_z\rangle$, normalized with respect to the corresponding homogeneous case vs the refractive index $n$. Each curve refers to a different size of
the system: $L=0.5\mu$m ($\bullet$), $1.0\mu$m ($\blacktriangle$) and $2.0\mu$m ($\blacksquare$). The coloured box marks the transition at which the OMF $z$-component is enhanced by the disorder.}\label{fig2}
\end{figure*}
%----------------end figure 2-----------------%
%--------------Figure 3-----------------------%
\begin{figure*}
\includegraphics[width=16cm]{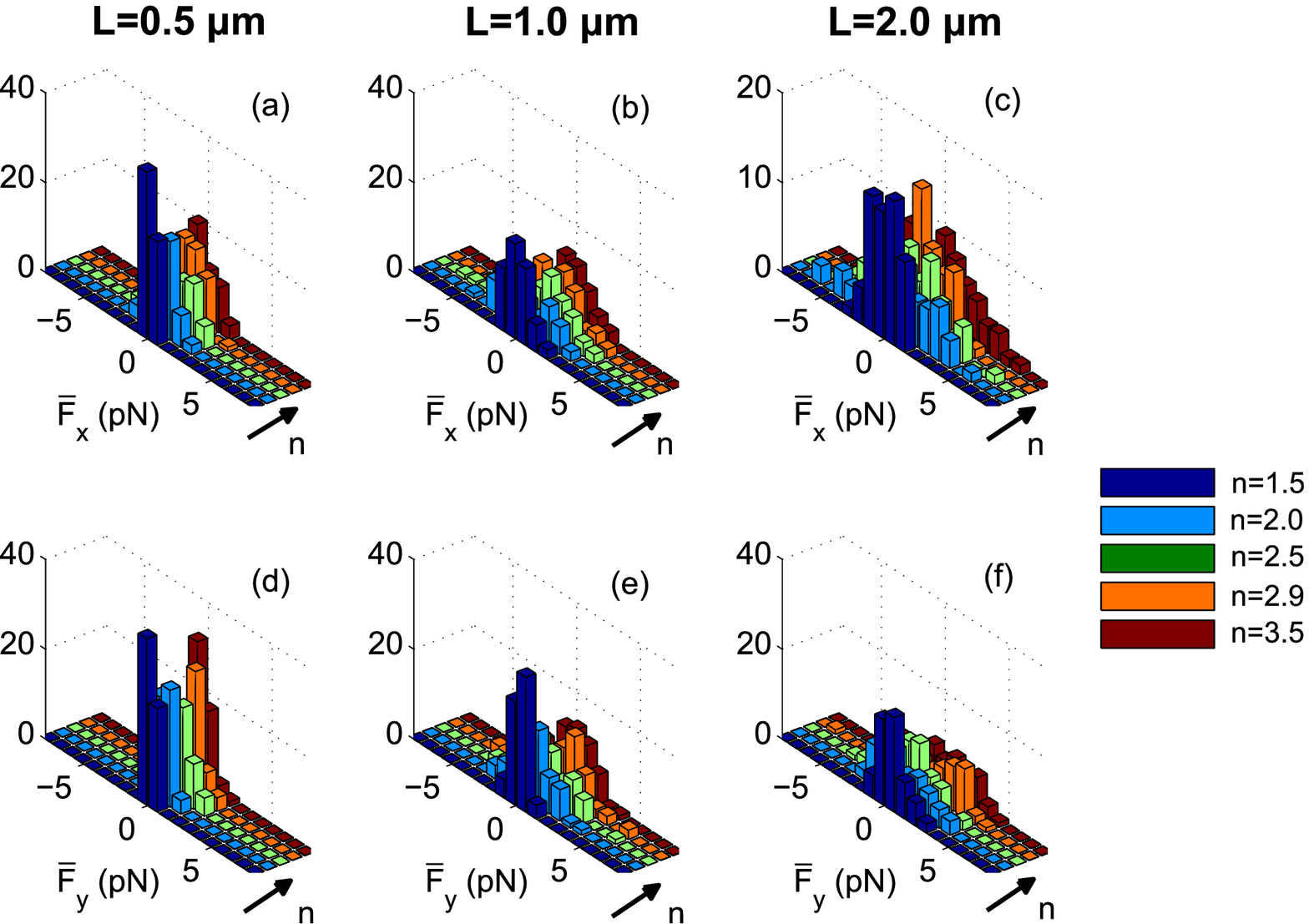}
\caption{\textbf{OMFs transverse components distribution.} Histograms of the stationary value of the transverse $(x,y)$ components of the optomechanical force, $\overline F_x$ and $\overline F_y$ as obtained for $60$
different disorders. Each panel corresponds to a given size of the system, $L=0.5\mu$m (a,d), $1.0\mu$m (b,e) and $2.0\mu$m (c,f)}\label{fig3}
\end{figure*}
%----------------end figure 3-----------------%

We consider a cubic structure with dimensions L$_x=$L$_y=$L$_z=$L, made by a random distribution of $100$nm radius monodispersed dielectric spheres with refractive index $n$. We consider three different sizes of the cubic box, L$=0.5$, $1.0$ and $2.0\mu$m; and we vary the refractive index of the single particle in an interval ranging from $n=1.5$ to $3.5$.
We solve the time-dependent Maxwell's equation in 3D spatial dimensions:
\begin{equation}
\begin{array}{l}
\nabla\times\mathbf{E}=-\mu_0\partial_t\mathbf{H}\\
\nabla\times\mathbf{H}=\partial_t \mathbf{D}\text{,}
\end{array}
\label{MaxwellEq}\end{equation}
where $\mathbf{D}$ is the displacement vector given by $\mathbf{D}=\epsilon_0\epsilon_r\mathbf{E}$, and $\epsilon_r=n^2$.
We make two series of simulations by launching two different x-polarized excitation signals on the $3$D structure: (i) we use a continuous wave (CW) light beam with wavelength $\lambda=600$nm to calculate the OMFs by the MSTM; (ii) to characterize the photons transport regime, we use a light pulse of duration t$_0=10$fs, with a spectral content centered at $\lambda=600$nm. Both the signals are launched along the $z$-direction and impinge on the input facet of the assembly located at $z=0$. The output facet is placed at $z=L$.

Let $\Sigma$ and $V$ be the surface area and the volume of the block, respectively, the time-dependent force due the EM wave, neglecting electrostriction, is given by \cite{brevik79}
\begin{equation}
\mathbf{F}=\frac{d \mathbf{G}_{mech}}{d t}=\int_\Sigma \overline{\overline{\mathbf S}}\cdot\mathbf{\hat n}\,dA-\frac{1}{c^2}\frac{d}{d t}\int_V \mathbf{E}\times\mathbf{H}\,dV
\label{stratton}
\end{equation}
where $\overline{\overline{\mathbf{S}}}\cdot\mathbf{\hat n}=\epsilon \mathbf{E}(\mathbf{E}\cdot \mathbf{\hat n})+\mu  \mathbf{H}(\mathbf{H}\cdot \mathbf{\hat n})- \small{\frac{1}{2}}(\epsilon E^2+\mu H^2) \mathbf{\hat n}$ is the projection of the Maxwell stress tensor $\overline{\overline{\mathbf{S}}}$ on the unitary normal $\mathbf{\hat n}$ exiting from the surface $\Sigma$.
The last term in (\ref{stratton}) is the time-derivative of the EM momentum in the volume $V$, whose density is $\mathbf{g}$ and given by the Abraham and von Laue expression $\mathbf{g}=\mathbf{g}_A=\small{\frac{1}{c^2}}\mathbf{E}\times\mathbf{H}$. \cite{Barnett:10}
In the CW case the time-average force is given by $\overline{\mathbf{F}}=\small{\frac{1}{T}}\int_{-T/2}^{T/2} \mathbf{F}dt$, with the optical cycle $T=\lambda/c$. $\overline {\bf F}$ gives the amount of momentum per unit time transferred to the block.
In the pulsed case, the time-average per single pulse is defined as
$\overline{\bf F}=\small{\frac{1}{T}}\int_{-\infty}^{\infty} \mathbf{F}dt$, which gives the total momentum transferred to the block per pulse during a normalization time $T$ much longer than the pulse duration.

We calculate the following output quantities: i) the three components of the resulting electromagnetic force, $\overline{\bf{F}}$, acting on the whole random assembly of dielectric beads; ii) the transverse [in the $(x,y)$ plane] intensity distribution of the electric field
at various $z$; iii) the total transmission $T(t)$ calculated  by integrating the $z$-component of the Poyinting vector over the output $(x,y)$ plane. Notice that the first two quantities are obtained by the CW simulations, while the latter quantity is obtained by the pulsed excitation.

\textbf{Optomechanical forces.} Figure \ref{fig1}a shows the sketch of the simulated system. The random assembly of dielectric spheres is illuminated by a $1\mu$m waist CW laser beam;  light scattered by the random structures in all the directions is also indicated; the image in panel (b) displays the typical intensity distribution of the intensity at the output plane ($z=L$). Figure \ref{fig1}c shows the time dynamics of the force $F_z(t)$ for L$=2\mu$m and $n=1.5$. The fast oscillations correspond to the optical carrier of the exciting source; the superimposed curve (thick line) is obtained by filtering out such oscillations. The time-dynamics of the filtered signal looks as a square-wave pulse due to the initial transient, which is needed by the EM wave to propagate through the whole structure. After this initial transient a stationary regime takes place. Notice that the stationary value of the longitudinal force is different from zero.

When the index contrast between the beads and the surrounding medium (vacuum) increases, the recoil force edge in the pulse is smoothed.
Figure \ref{fig1}d shows the effect of the scattering on the time behavior of $\overline F_z(t)$ for a specific disorder realization. As anticipated, when $n$ grows, the trajectory of the photons follows a complex path, which affects $\overline F_z(t)$.

Figure \ref{fig1}e shows the temporal behavior of the filtered transverse components $\overline F_{x,y}(t)$.
Because of the random walk due to the multiple scattering, the photons escaping from the lateral sides of the sample
generate transverse OMFs components.
When averaging over several disorder realizations, these components vanish; this transverse random force
is expected to be observable for a specific sample with a fixed disorder realization.
In order to determine the statistical distribution of the OMFs, we repeated the CW simulations for sixty different disorder realizations for any
considered system size $L$ and refractive index $n$.

The lateral leakages of photons increase with the scattering strength; this reduces the recoil force in the $z$-direction and increases the overall longitudinal force. Such a dynamics is confirmed by the behavior of the disorder average optical force $z$-component $\langle\overline F_z(t)\rangle$ reported in Fig. \ref{fig2}a-c for different $L$ and $n$.
Figure \ref{fig2}d shows the stationary value of the $\langle\overline F_z(t)\rangle$ curves in Fig.\ref{fig2}a-c versus $n$, normalized with respect to the correspondent homogeneous structure (i.e., a cubic block of the same size of the random assembly and refractive index).
Notice that, for the largest sample (L$=2.0\mu$m), one can identify a refractive index threshold, at which the longitudinal force in the disorder case is enhanced with respect to the homogeneous case.

We then analyse the transverse components, $\overline F_x(t)$ and $\overline F_y(t)$. Figures \ref{fig3} a-d show the statistical distributions of the stationary values of $\overline F_x$ and $\overline F_y$. Notice that the histograms broaden with the increase of the refractive index and that this effect is more evident for the largest sizes. This analysis predicts that a specific disorder realization sustains a transverse force at a random direction.

%--------------Figure 4------------------------%
\begin{figure*}
\includegraphics[width=16cm]{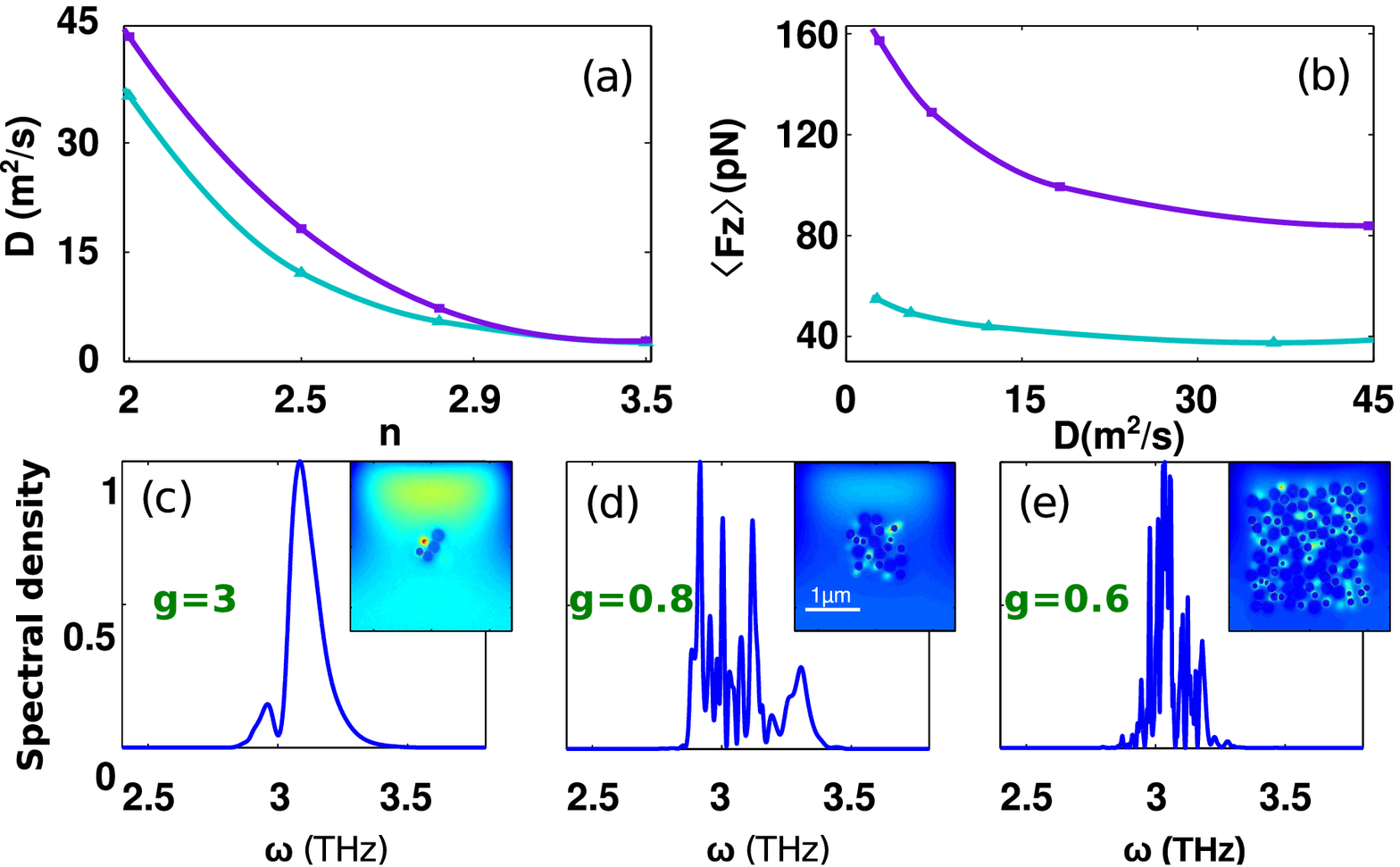}
\caption{\textbf{Anderson localization and optomechanical forces.} (a) Diffusion coefficient $D$ vs $n$ obtained for the system sizes L$=1.5\mu$m ($\blacksquare$) and $2.0\mu$m ($\blacktriangle$).
(b) Stationary value of the $\langle\overline F_z\rangle$ vs diffusion coefficient D for L$=1.5$ ($\blacktriangle$) and $2.0\mu$m ($\blacksquare$) samples. (c-e) Spectral content of the electric field at the output face of sample
with the highest particles refractive index ($n=3.5$) and different length L$=0.5$ (c), $1.0$ (d) and $2.0\mu$m (e). The dimensionless conductivity $g$ is also indicated. The insets show the
intensity spatial distributions of the electric fields in the $(x,y)$ plane.}\label{fig4}
\end{figure*}
%----------------end figure 4-----------------%
\textbf{Optomechanics at onset of the Anderson localization. }
In order to link the OMFs with the photon dynamics, i.e., with the transition towards a localized regime expected at large $n$, we perform a set of simulations by considering an input EM pulse with duration $t_0=10$~fs  and central wavelength $\lambda=600$nm.
Following the procedure in \cite{Gentilini09}, we analyse the trailing edge of the total transmission $T(t)$, collected at the output facet of the system, and determine the diffusion coefficient as $D=L^2/(\pi^2\tau)$, with $\tau$ the exponential decay time of $T(t)$.

Figure \ref{fig4}a shows the diffusion constant $D$ versus $n$ for the $L=1.0\mu$m (circles) and $L=2.0\mu$m (triangles). Figure \ref{fig4}b shows the disorder average OMFs $z$-component versus $D$. An enhancement of the OMFs is found when the diffusion constant reduces. The OMFs reach a maximum when $D$ reaches the smallest value in correspondence of the localization transition.
This result clearly shows the relation between the OMFs and the photons dynamics within a disordered 3D medium.
Notice that in order to calculate the forces, one needs to consider a spatially limited structure for which the diffusion
constant does not vanish because of the finite size effects.

In order to further characterize the photon transport regime, following the scaling theory of localization in 3D systems \cite{ShengBook},
we analyse the spectral content of the output electric field when varying the system size. Figures \ref{fig4} (c-e) report the output EM spectra corresponding to samples with $n=3.5$ and L$=0.5$, $1.0$ and $2.0\mu$m. When increasing L, the spectra display an increasing number of peaks. The linewidth of these resonances narrows with L, signaling the formation of long living localized modes.
A measure of the localization degree is given by the Thouless conductivity $$g(L)=\frac{\delta\omega}{\Delta\omega},$$ where $\delta\omega$ is the spectral peak width and $\Delta\omega$ the peak distance, averaged over all the modes. The localization transition occurs when $g<1$ \cite{ShengBook}, a condition that is found in our numerical experiments when $n=3.5$ and L=$1.0$ and $2.0\mu$m. For these values, the output electric field intensity $(y,z)$ profile is reported in Fig.~\ref{fig4}c-e. We remark that $g>1$ for all the considered case with $n<3.5$. This result shows that for $g<1$ the force reaches a maximum. {\it In other words, the onset of the Anderson localization has an optomechanical counterpart.}

{\it Conclusions.}
To unveil the effect of disorder on the optomechanical forces acting on a three-dimensional random assembly of dielectric particles,
we have used a massively parallel computational approach.
The study of the time behavior of the optical pressure and of the statistics of the transverse components of the optical force, combined with the analysis of the light transport regime, has allowed to quantitatively establish a relation between the forces and the photons diffusion coefficient $D$.
We have found that the momentum transferred to a disordered micron-sized object increases when light approaches a localized regime. The onset of the Anderson localization enhances the mechanical action of light.
These findings may open the road to the exploitation of light scattering and localization for the motion of complex dielectric structures by means of low power laser sources. A proper arrangement of the refractive index distribution, loss, shape, size, and spatial configuration of dielectric particles, allows to control the photon Brownian motion and, correspondingly, to engineer the light induced forces, this makes possible the realization of optically activated  micro-motors suitable for an enormous amount of applications.

\emph{Acknowledgments-}Our parallel code is a C++ 3D+1 FDTD based on the MPI-II protocol and running on the FERMI IBM Blue Gene Q system at CINECA, within the Italian Super Computing Resource Allocation (ISCRA) initiative.

\end{document}